\documentclass[%
 reprint,
 amsmath,amssymb,
 aps
]{revtex4-2}

\usepackage{graphicx}
\usepackage{lipsum}
\usepackage{graphicx}
\usepackage{dcolumn}
\usepackage{bm}
\usepackage{hyperref}

\begin{document}
\title{Spontaneous Symmetry Breaking of an Optical 
Polarization State in a Polarization-Selective Nonlinear Resonator}

\author{K. S. Manannikov}
 \altaffiliation{Corresponding author: konstantin.manannikov@weizmann.ac.il}
\affiliation{Department of Physics of Complex Systems, Weizmann Institute of Science, Israel}%
\author{E. I. Mironova}%
\affiliation{Department of Physics of Complex Systems, Weizmann Institute of Science, Israel}%
 \author{A. S. Poliakov}
\affiliation{Department of Physics of Complex Systems, Weizmann Institute of Science, Israel}%
\author{A. Mikhaylov}
\affiliation{Q.ANT GmbH, Handwerkstr. 29, 70565 Stuttgart, Germany}
\author{A. E. Ulanov}%
\affiliation{Deutsches Elektronen-Synchrotron DESY, Notkestr.~85, 22607 Hamburg, Germany}%
\author{A. I. Lvovsky}%
\affiliation{Clarendon Laboratory, University of Oxford, Oxford OX1 3PU, UK
}%

\begin{abstract}
We exploit polarization self-rotation in atomic rubidium vapor to observe spontaneous symmetry breaking and bistability of polarization patterns. We pump the vapor cell with horizontally polarized light while the vertical polarization, which is initially in the vacuum state, is resonated in a ring cavity. Microscopic field fluctuations in this mode experience cumulative gain due to the compound action of amplification due to the self-rotation and feedback through the resonator, eventually acquiring a macroscopic magnitude akin to an optical parametric oscillator. The randomness of these fluctuations results in a bistable, random macroscopic polarization pattern at the output. We propose utilizing this mechanism to simulate Ising-like interaction between multiple spatial modes and as a basis for a fully optical coherent Ising machine.
\end{abstract}

\maketitle

\begin{figure}[t!]
\center{\includegraphics[width =0.99\linewidth]{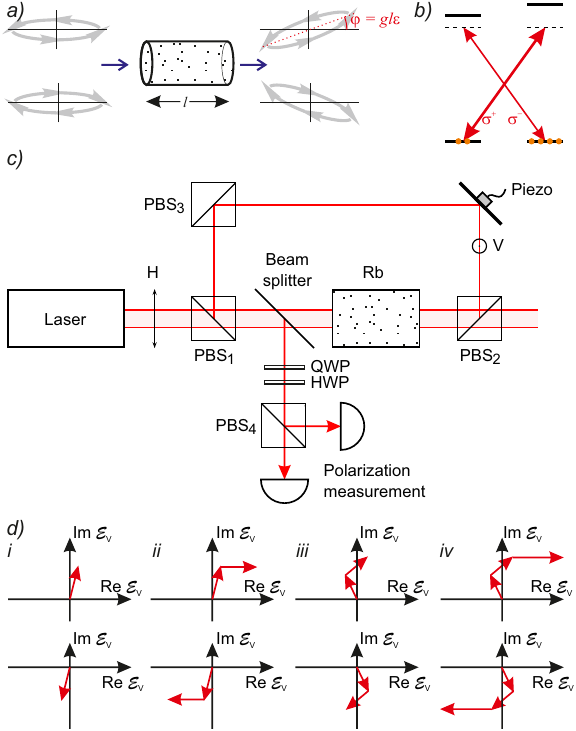}}
\caption{Concept of the study. a) PSR effect. The magnitude and direction of self-rotation depend on the ellipticity of the input polarization and its handedness, respectively. b) Conceptual explanation in an atomic X-system. Elliptically polarized light is described as a sum of two circular components  ($\sigma_{+}$ and $\sigma_{-}$) with unequal intensities, interacting with the respective transitions in the X-system. These components experience different refractive indices due to optical pumping and ac Stark shift. c) Scheme of the experiment. Rubidium vapor is pumped by horizontally polarized light, while the vertical polarization is resonated in a ring cavity. PBS: polarizing beam splitter; HWP: half-wave plate; QWP: quarter-wave plate. d) Origin of optical parametric oscillation and bistability. Complex electric field vector of the vertical polarization component is shown: (i) before entering the vapor cell, (ii) after passing through the cell and experiencing PSR, (iii) after a roundtrip through the cavity, before entering the cell for the second time, (iv) after second passage through the cell. The phase of the developed oscillation depends on the sign of the initial ${\rm Im}\mathcal{E}_V$, giving rise to bistability (see text for further details).}
\label{concept}
\end{figure}

\paragraph{Introduction.}
A nonlinear optical effect known as polarization self-rotation (PSR)  occurs as a result of the interaction of an elliptically polarized light with a nonlinear $\chi^{(3)}$ medium. PSR consists in the rotation of the polarization ellipse at a rate that increases with the ellipticity [Fig.~\ref{concept}(a)]. In a Kerr medium, the refractive index for each component depends on its intensity. Unequal refractive indices result in different phase velocities, which in turn leads to a rotation of the polarization ellipse [Fig.~\ref{concept}(b)]. 

Initially investigated in the 1970s \cite{maker1964intensity}, PSR gained renewed attention in the early 2000s, when it was considered as a simple source of squeezed light \cite{matsko2002vacuum, ries2003experimental, mikhailov2008low}. When a PSR-exhibiting medium is pumped with linearly (e.g., horizontally) polarized light, microscopic field fluctuations in the orthogonal (vertical) polarization mode can manifest as microscopic ellipticity of the overall polarization pattern. Microscopic self-rotation due to this ellipticity brings about linear shear of the phase space of the vertical polarization mode, squeezing the uncertainty circle. 

In this work, we experimentally investigate the optical parametric oscillation (OPO) induced by PSR. To that end, the effect on the vertical mode is enhanced by a cavity in which that mode resonates [Fig.~\ref{concept}(c)]. For certain resonator lengths, the amplification becomes cumulative, leading to a macroscopic electromagnetic field amplitude in the steady state and thus to a macroscopic ellipticity of the field polarization in the vapor cell. 

Our PSR-based OPO shows a spontaneous symmetry breaking of the optical polarization state. The handedness of the elliptical pattern in the steady-state is determined by the sign of the initial  fluctuation from which it originated. Consequently, this handedness is random every time the parametric oscillation is initiated and remains constant for as long as the oscillation is sustained. In other words, the polarization state is bistable.

Bistability is a common feature of parametric oscillators 
and Kerr cavities, observed in a variety of schemes \cite{zheludev1989polarization, gauthier1990instability, gauthier1990bistability, coen2024topologicalbist}. 
For example, Moroney {\it et al.} \cite{moroney2022kerr} observed polarization bistability with a continuous wave in a high-Q fiber resonator with Kerr nonlinearity. Due to spontaneous symmetry breaking, one circular component of the linearly polarized input light became dominant inside the cavity, while the other one was reflected. 

Bistable systems can be  affected by various asymmetries introducing a bias in the final binary distribution \cite{moroney2022kerr,steinle2017unbiased,
quinn2023random, PhysRevResearch.2.023244}. In our scheme, the main source of the signal's polarization bias is the leakage of the pump in the cavity feedback loop. 
However, polarization selectivity of the resonator in combination with the full control of the pump polarization  allow us to minimize that leak and observe random selection of one of the two helicities in the output.

Atomic resonant enhancement in our system leads to a much stronger nonlinearity than the $\chi^{(3)}$ nonlinearity in a fiber or the $\chi^{(2)}$ nonlinearity in a crystal: significant squeezing of a continuous wave mode can be obtained in a single pass through the vapor cell. This makes the scheme attractive for the implementation of an all-optical coherent Ising machine. We discuss this idea at the end of the paper.

\paragraph{Concept}

We analyze the transformation of the field polarization pattern under the joint action of  PSR and the resonator. Interacting with a rubidium vapor cell, elliptically polarized light experiences rotation of its optical axis by the angle $\varphi(\varepsilon)$, where $\varepsilon$ is the ellipticity. 
Subsequently, the vertical polarization is reflected by a polarizing beam splitter into the resonator [Fig.~\ref{concept}(c)]. During a roundtrip through the resonator, the field accrues a phase shift $\psi$ and experiences a linear loss before being re-injected into rubidium. The field transformation in one roundtrip can thus be written as follows:
\begin{align}\label{reseq}&
\begin{pmatrix}
  \mathcal{E}_H\\
  \mathcal{E}_V^{(n+1)}
\end{pmatrix}
\\ \nonumber &\hspace{1em}=
\begin{pmatrix}
  0 & 0\\
  0 & \sqrt\eta e^{i \psi}
\end{pmatrix}
\begin{pmatrix}
  \cos(\varphi) & -\sin(\varphi)\\
  \sin(\varphi) & \cos(\varphi)
\end{pmatrix}
\begin{pmatrix}
  \mathcal{E}_H\\
  \mathcal{E}_V^{(n)}
\end{pmatrix}+\begin{pmatrix}
  \mathcal{E}_H\\
  0
\end{pmatrix},
\end{align} 
where  $\mathcal{E}_H$, $\mathcal{E}_V$ are the complex field amplitudes of the polarization components in the $n$th roundtrip and $1-\eta$ is the intensity loss in the cavity per roundtrip. 

Fig.~\ref{concept}(d) visualizes the evolution of $\mathcal{E}_V$ in the complex plane. We use the horizontal component as phase reference, so $\mathcal{E}_H$ is a real number. PSR admixes this real horizontal amplitude into the vertical mode
[Fig.~\ref{concept}(d,ii)]. The subsequent roundtrip through the resonator rotates the complex phase of the field and scales it by factor $\sqrt\eta$ [Fig.~\ref{concept}(d,iii)]. The resulting vector undergoes further horizontal displacement due to self-rotation (determined by the new ellipticity) [Fig.~\ref{concept}(d,iv)], and so on. 

To identify the conditions under which this recurrent transformation leads to oscillation, we assume the initial ellipticity to be small ($\varepsilon\ll 1$). The self-rotation angle is then proportional to the ellipticity  \cite{matsko2002vacuum}:
\begin{equation}
   \varphi \approx g l \varepsilon,
   \label{eq:phi}
\end{equation} 
where the proportionality coefficient $g$ depends on the intensity of the incident light and its detuning from resonance. The ellipticity itself (for $\mathcal{E}_H\in \mathbb{R}$) is defined as $\epsilon=\arcsin\dfrac{\mathcal{E}_H{\rm Im} \mathcal{E}_V}{|\mathcal{E}_H|^2+|\mathcal{E}_V|^2}$ \cite{matsko2002vacuum}, simplifying to $\varepsilon\approx\dfrac{{\rm Im} \mathcal{E}_V}{\mathcal{E}_H}$ for $\varepsilon \ll 1$. The vertical field component transformation according to Eq.~\ref{reseq} then becomes 

\begin{equation}\label{transPSR1}    \mathcal{E}_V^{(n+1)}
\approx\sqrt\eta e^{i \psi}(\mathcal{E}_V+gl{\rm Im \mathcal{E}_V}).
\end{equation}
The threshold condition corresponds to $\left|\mathcal{E}_V^{(n+1)}\right|>\left|\mathcal{E}_V^{(n)}\right|$.
This inequality can be solved by writing Eq.~\ref{transPSR1} in the matrix form with respect to the real and imaginary parts of $\mathcal{E}_V$ and finding the eigenvalues $\lambda_{1, 2}$ of the transfer matrix. The threshold condition is obtained by requiring $\max_{{\psi}}(|\lambda_{1, 2}|) > 1$, which resolves to
\begin{equation}
gl > 1 / \sqrt{\eta} - \sqrt{\eta},
\label{threshold}
\end{equation}
with the highest gain observed at
\begin{equation}
\psi = \arctan\left(\dfrac{gl}{2}\right).  
\end{equation}
The infinite growth of $\left|\mathcal{E}_V\right|$ is limited by the pump depletion and saturation of the self-rotation coefficient for large ellipticities.

To obtain the polarization of the field in the steady state, the small-ellipticity condition must be relaxed, and Eq.~\ref{reseq} must be solved numerically.

The bistability of this PSR-based OPO can be understood by examining the top and bottom rows of Fig.~\ref{concept}(d). As is evident from Eq.~\ref{transPSR1}, the sign of the imaginary component of $\mathcal{E}_V$ (and hence the helicity of the polarization pattern) is preserved in each roundtrip through the resonator. Depending on the initial value of this sign, $\mathcal{E}_V$ will acquire one of the two opposite phases when amplified. Because the self-rotation angle $\varphi(\varepsilon)$ is an odd function of the ellipticity, it follows that whenever a vector $(\mathcal{E}_H,\mathcal{E}_V)$ with ellipticity $\varepsilon$ is a steady state solution of Eq.~\ref{reseq}, so is the vector $( \mathcal{E}_H, \mathcal{E}^*_V)$ with ellipticity $-\varepsilon$.

\paragraph{Experiment}
\begin{figure}[h!]
\center{\includegraphics[width =\linewidth]{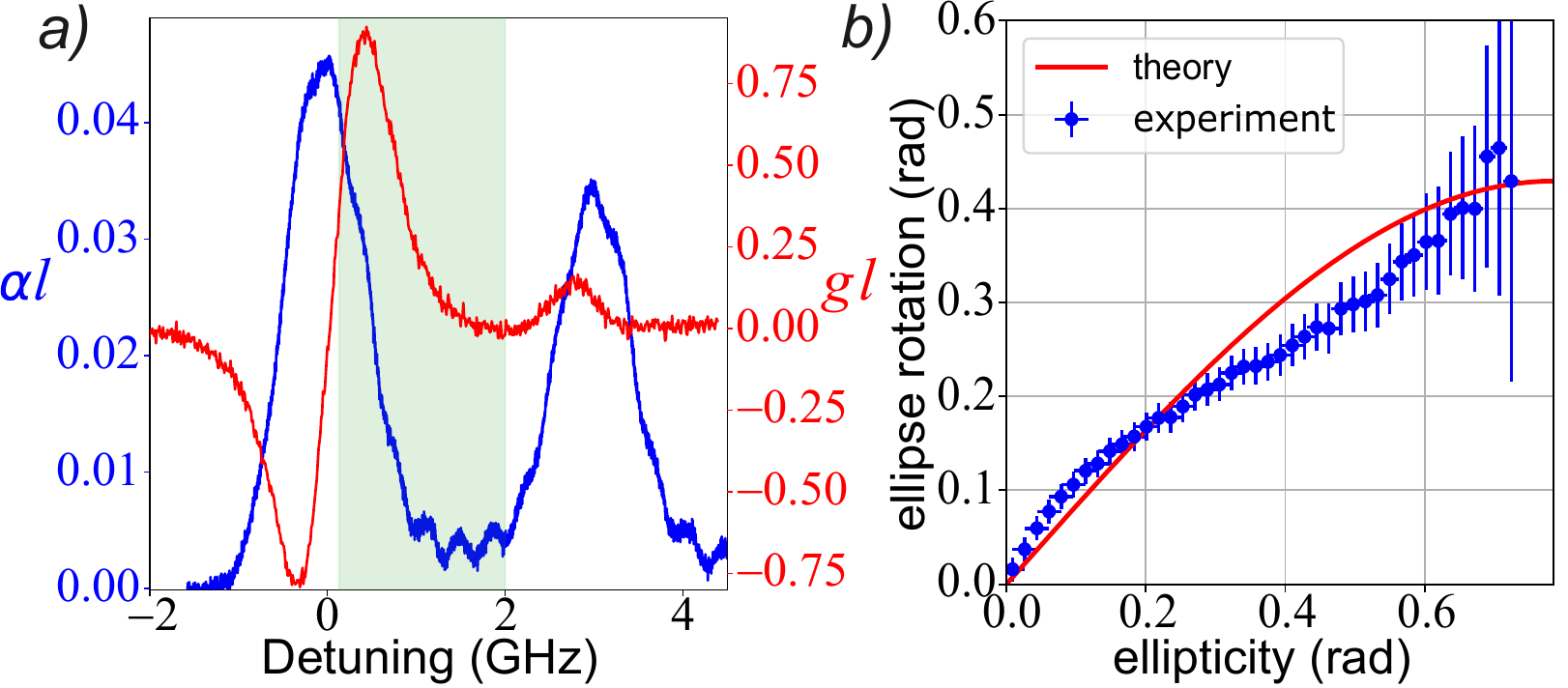}}
\caption{Experimental observation of PSR. Data for both plots are acquired at the pump power $P = 13.5$ mW. a) Coefficient of absorption $\alpha l$ (blue line, left scale) and self-rotation parameter $gl$ (red line, right scale) measured across the ${ }^{85} \mathrm{Rb}$ D1 transition. Zero detuning corresponds to the maximum absorption wavelength. The absorption coefficient $\alpha l$ is calculated as the natural logarithm of the ratio of the transmitted and input intensities. The green region marks the frequency range where bistability is observed. b) Dependence of the polarization ellipse rotation angle on initial ellipticity at frequency detuning $\Delta = 0.35$ GHz. Experimental data (blue dots) is compared to the theoretically obtained curve (red line) from Eq.~\ref{phiepsilon}.
}
\label{fig:phi_eps}
\end{figure}

For the experiment, we utilize a Vitawave external-cavity diode laser, which emits continuous-wave light at 795 nm with a power of 13.5 mW, resonant with the $D_1$ transition in $^{85}\mathrm{Rb}$. For PSR measurements, the laser is set to horizontal polarization and directed into a rubidium cell heated to $70^\circ$C and placed inside a telescope formed by two $f=30$ cm lenses. The transmitted beam is subjected to polarimetric balanced detection in a $45^\circ$ basis (see Supplementary 1 for more experimental details). 

The self-rotation is measured as a function of the laser detuning [Fig.~\ref{fig:phi_eps}(a)] and ellipticity [Fig.~\ref{fig:phi_eps}(b)]. For the latter dependence, we also obtain a theoretical prediction based on an X-shaped level structure [Fig.~\ref{concept}(b)] (more details in Supplementary 2):
\begin{equation}
\begin{gathered}\label{phiepsilon}
     \varphi = 
     \dfrac{C \delta \sin(2 \varepsilon)}{ (2 + 8 \delta^2) + [1 +  \cos(4 \varepsilon)]I/I_{\rm sat}},
\end{gathered}
\end{equation}
where $I$ is the optical intensity, $I_{\rm sat}$ the saturation intensity of the transition, $\delta$ the ratio of the detuning and the resonance line width, and $C$ a proportionality coefficient. A good fit to the experimental data is found for $\delta=0.1$ and $I/I_{\rm sat}=10$. We can see that the proportionality $\varphi\propto\varepsilon$ holds for small ellipticities. The difference between the experimental and theoretical curves is due to simplified treatment of the complex rubidium energy structure as a four-level system without Doppler broadening.

\begin{figure}[h!]
\includegraphics[width=\linewidth]{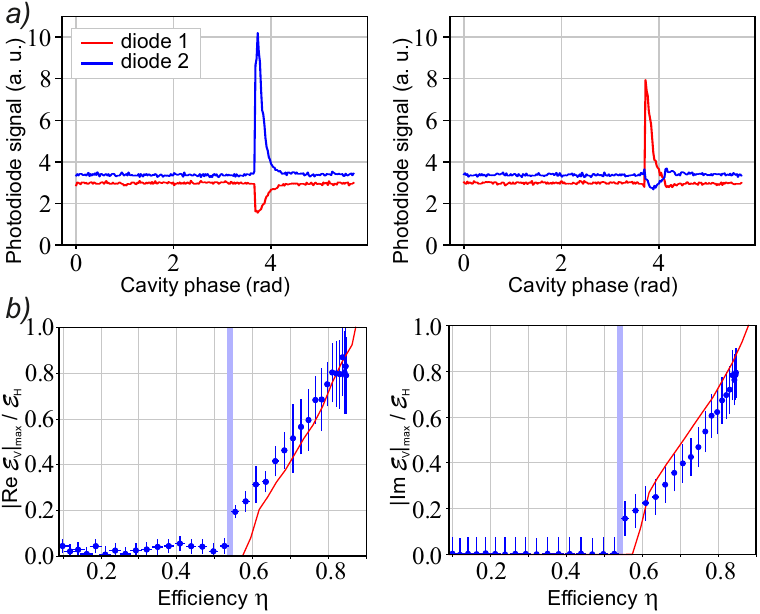}
\caption{Bistability measurements. a) Signals from the two photodiodes performing polarization measurements in the circular basis during two successive phase scans of the cavity.  b) Real and imaginary parts of the steady-state vertical amplitude as a function of loss in the resonator. The red theoretical curve is obtained by numerically solving Eq.~\ref{reseq} with the self-rotation angle computed from Eq.~\ref{phiepsilon}. The blue shaded area shows the threshold calculated from Eq.~\ref{threshold} with the self-rotation parameter $g$ obtained from the data in Fig.~\ref{fig:phi_eps}(b). All measurements are performed with a fixed frequency detuning $\Delta = 0.35$ GHz and pump power $P = 13.5$ mW. Small non-zero values of  ${\rm Re}\mathcal{E}_V$ under the threshold is due to imperfect balancing of the polarization measurement.}
\label{los_and_ell}
\end{figure}

To observe bistability, we build the setup shown in Fig.~\ref{concept}(c) (see Supplementary 4 for details). The laser is tuned into the range with the lowest losses and the highest nonlinearity [shown in pale green in Fig.~\ref{fig:phi_eps}(a)]. A small fraction of the field is deflected from the cavity before the cell with a highly transmissive beam splitter, which was tested to have a negligible effect on the relative phase of the horizontal and vertical polarization. The reflected beam is subjected to polarimetric measurement in the circular basis via two photodetectors. The pump polarization is carefully adjusted by two waveplates to eliminate leakage into the cavity
in the absence of PSR.

The signal from these photodetectors, while the phase of the resonator is scanned with a piezoelectric transducer, is shown in Fig.~\ref{los_and_ell}(a). Whenever the phase passes through a cavity resonance, oscillation emerges with randomly either the right or left circular component being prevalent. 

To investigate the dependence of the bistable state on the losses in the resonator, we measure the output polarization in circular and canonical bases to evaluate (see Supplementary 3 for a derivation):
\begin{equation}
\dfrac{{\rm Im}\mathcal{E}_V}{\mathcal{E}_H} = \dfrac{I_L - I_R}{2 I_H} \textrm{ and }\dfrac{{\rm Re}\mathcal{E}_V}{\mathcal{E}_H} = \dfrac{\sqrt{I_V - ({\rm Im}\mathcal{E}_{V})^2}}{\mathcal{E}_H},
\label{Pvloop}
\end{equation}
where $I_{R}$, $I_{L}$, and $I_H$ are the intensities of the right circular, left circular, and horizontal polarization components respectively. The resulting data are demonstrated in Fig.~\ref{los_and_ell}(b). We observe that the bistability is significantly tolerant to losses, a consequence of a high phase-dependent gain in each roundtrip.

To demonstrate the randomness of the bistable states, we acquire a series of $M=700$ oscillation events, ascribe a value $s_n=\pm1$ to each event according to its helicity, and calculate the auto-correlation
\begin{equation}
    K(m) = \dfrac{1}{M} \sum\limits_{n=0}^M s_{n} \cdot s_{n + m}.
\end{equation}

\begin{figure}[h!]
\center{
  \includegraphics[width = 0.75\linewidth]{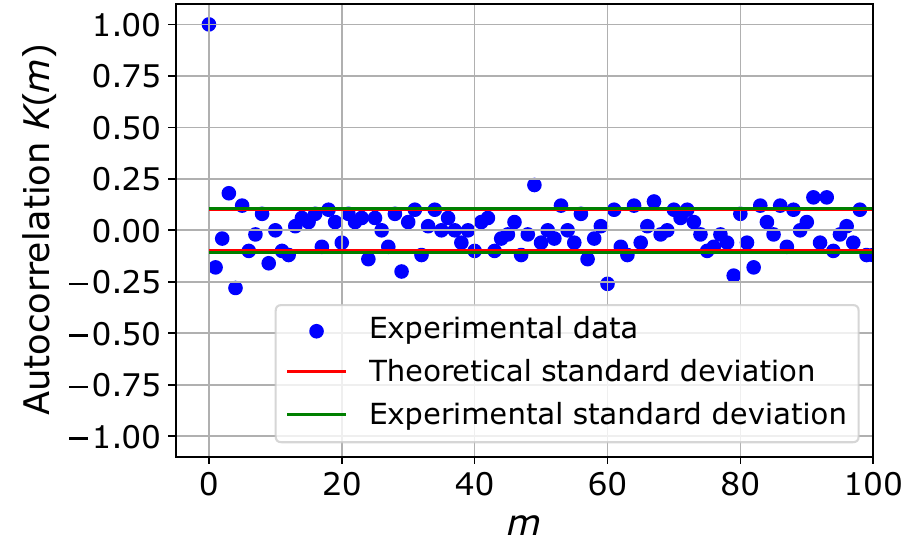} }
 \caption{Auto-correlation function for experimental values of binary phases (blue dots). Double standard deviation intervals for the experimental auto-correlation function and for the auto-correlation function of the ideal independent binary distribution are also shown.}
\label{figureautocorr_spins}
\end{figure}

As can be seen in Fig.~\ref{figureautocorr_spins}, this function behaves similarly to that of a fair coin: the standard deviations of $K(m)$ (for $m\ne0$) for the experimental data and ideal Bernoulli distributions are almost equal. This suggests that the randomness observed is likely to originate from quantum vacuum fluctuations, as is normally the case with OPOs \cite{steinle2017unbiased}. However, it cannot be excluded that the fluctuations are of classical origin, such as fast oscillations of the pump polarization or external magnetic field.

Such randomness can be a basis for a random number generator \cite{steinle2017unbiased,
quinn2023random}. While the random data in this experiment are acquired at a rate limited by the cavity scan speed, a more fundamental physical limitation is imposed by the atomic relaxation time and the duration of several roundtrips through the resonator, both on time scales of tens of nanoseconds. Higher rates can be obtained by spatial demultiplexing, as discussed below.

\paragraph{Summary and outlook}
In summary, we demonstrated that polarization self-rotation could be utilized to generate bistable polarization states of light. Spontaneous symmetry breaking for the initially horizontally polarized light was achieved thanks to the resonant optical nonlinearity of rubidium vapor cell in a vertical polarization selective resonator. As a result, the field in the resonator becomes elliptically polarized with a random helicity, as evidenced by statistical analysis. 

The optical system presented in this paper can be scaled up: several independent bistable spatial modes can be obtained simultaneously by focusing multiple spatially separated laser beams in the same vapor cell. These modes can be arbitrarily coupled by placing a spatial light modulator into the resonator using spatial matrix-vector multiplication methods developed in the context of optical neural networks \cite{spall2022hybrid,bernstein2023single}. This system of coupled optical parametric oscillators would implement an all-optical coherent Ising machine (CIM), i.e.~an analog optical neural network capable of evolving into the ground state of an Ising Hamiltonian \cite{strinati2020,strinati2021}. 

In the original optoelectronic CIM  \cite{Inagaki2016,Mcmahon2016,honjo2021}, the pulsed modes are coupled through classical measurement and feedback, which precludes their entanglement and hence quantum computational advantage \cite{SimCIM}. An all-optical CIM, in which the interaction between modes occurs via interference, would address this shortcoming. Existing solutions, based on pulsed modes and fiber interferometers, either involved very few oscillators \cite{Marandi2014} or had the coupling limited to nearest neighbors \cite{Inagaki2016a}. Leveraging the resolution of spatial light modulators to couple spatially separated continuous-wave bistable modes
appears to be a promising path toward scalability.  The number of modes in this setting is then limited by the vapor cell area. While each mode in the cell is $\lesssim 0.25$ mm wide, spurious cross-talk between modes may impose a lower limit on the distance between neighboring modes. This effect can be curtailed by constructing an array of microcells with longitudinal partitioning. 

In addition to applications for combinatorial optimization, it is interesting to explore the potential of this all-optical CIM  as a quantum simulator of condensed matter physics. Such studies have been actively pursued with the D-Wave annealer, such as e.g.~a recent simulation of nonequilibrium dynamics of a magnetic spin system undergoing a quantum phase transition~\cite{king2024}. The promise of multiple coupled OPOs in this context has been shown theoretically \cite{Strinati2019,bello2019}, but experimental research has so far been limited to the optoelectronic scheme \cite{takesue2023}.

The bistability, especially near the exceptional point, is highly sensitive to fluctuations of the system parameters, particularly an external magnetic field. Hence the setup may also prove promising as a magnetic field sensor. 

We believe the system presented in this work offers significant potential for many academic and practical applications in quantum optics and analog computing. 

\paragraph*{Acknowledgments}

We are grateful to A.~V.~Masalov for support in the experiment and fruitful discussions.
AL's research is supported by the Innovate UK Smart Grant
10043476 and the EPSRC Grant EP/Y020596/1.

\paragraph*{Data Availability Statement} The data presented here are available from the corresponding author upon request.

\bibliography{main}

\newpage

\pagebreak

\section*{Supplementary}

\section{Self rotation: experiment}
\label{self-rot exp}
\label{expselfrot}
\begin{figure}[h]

\center{\includegraphics[width = 8.6 cm]{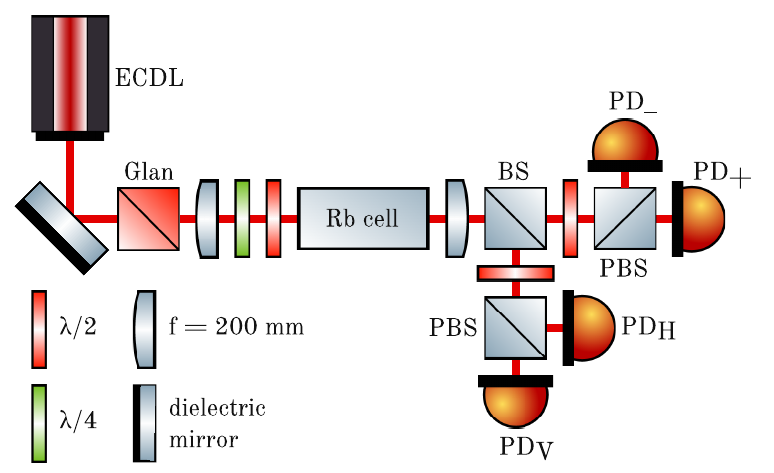}}
	\caption{Experimental setup for measuring PSR.}
	\label{ris:sr-scheme}
\end{figure}

A scheme of the experimental setup for measuring PSR is shown in Fig.~\ref{ris:sr-scheme}. The beam from an external cavity-diode laser (Vitawave ECDL-7950R) delivers approximately $50 \mathrm{~mW}$ of optical power in a single longitudinal mode. The laser is tuned across the rubidium D1 line ($\lambda \simeq 795 \mathrm{~nm}$). The beam is shaped by a pair of lenses and then transmitted through a Glan prism that transmits strictly linear polarization. A wave plate is placed before the Glan prism to vary the pump intensity; the transmitted intensity is 13.5 mW unless stated otherwise. The horizontally polarised light beam is then focused by a lens into a cylindrical vapor cell ($25 \mathrm{~mm}$ in diameter and $75 \mathrm{~mm}$ in length) with antireflection coated windows, filled with ${ }^{85} \mathrm{Rb}$ without buffer gas at 70$^\circ$C. With an estimated focal beam diameter of $d = 50 \,\mu\mathrm{m}$, we experimentally obtain $I_{\rm sat} \simeq 0.8 \mathrm{~mW}/ \pi d^2$, thus operating in a high saturation regime. 

To produce and control ellipticity in the initially horizontally polarised beam, a fine-adjustable $\lambda / 4$ plate is placed in front of the cell. To compensate for the impact of birefringent windows on pumping light we used a procedure described in Section 4.

\begin{figure}[ht!]
\includegraphics[width = 0.99 \linewidth]{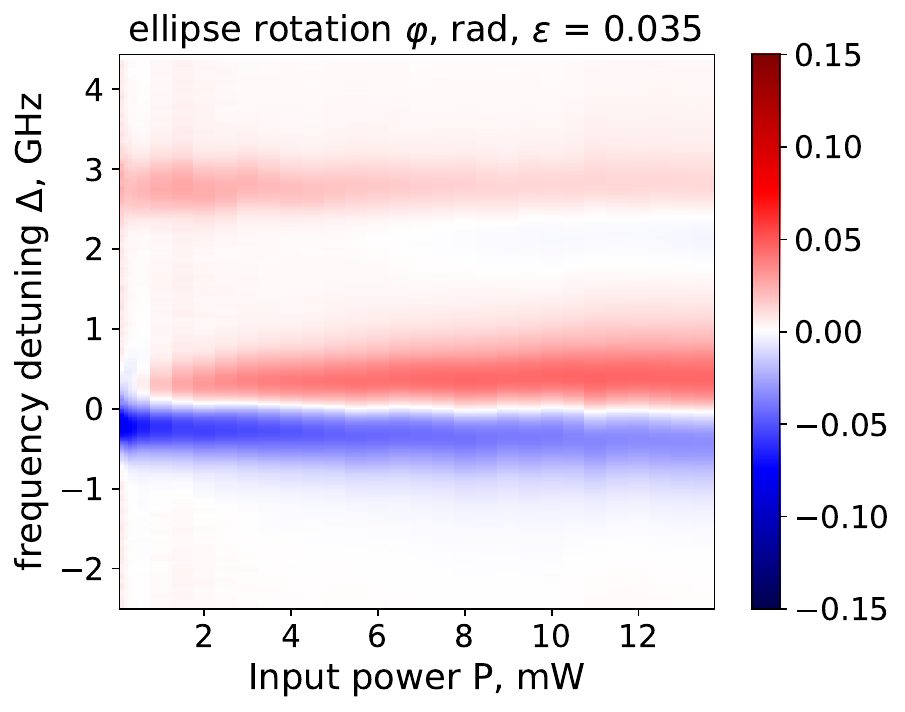}
\includegraphics[width = 0.99 \linewidth]{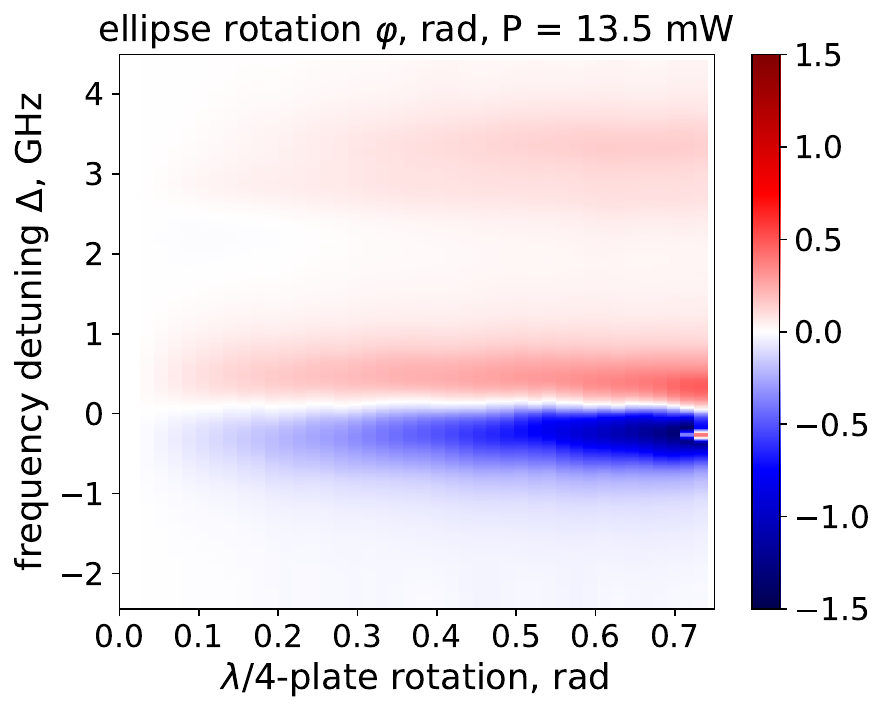}
\caption{Spectrum of the PSR angle $\varphi$.  (a) Ellipticity $\varepsilon = 0.035$ fixed, optical power variable.  (b) Optical power $P = 13.5$ mW fixed, ellipticity variable.}
\label{fig:color}
\end{figure}

After exiting the cell, the beam is directed to the measuring system consisting of a $ \lambda /2$ wave plate and a PBS, whose axis is aligned with the Glan prism. By rotating this $ \lambda /2$ plate, we can measure the signal in either a canonical or diagonal basis. The outputs of the PBS are directed to two photodiodes. Data are gathered by tuning the laser through the rubidium D1 line for a range of initial ellipticities and recording the signal from the photodiodes. The rotation angle of the polarisation ellipse is determined by (see Supplementary 3)
\begin{equation}
    \varphi = \dfrac{1}{2}\mathrm{arctan}\left(\dfrac{I_{+} - I_{-}}{I_{+} + I_{-} - 2I_{V}}\right),
\label{eq:selfrot_formula}
\end{equation}
where $I_\pm$ and $I_V$ are the intensities of the $\pm45^\circ$ and vertical polarizations, respectively. The measurements were performed with a small initial ellipticity $\varepsilon$ and the input intensity significantly below saturation. The data obtained under these conditions are shown in Fig.~2 of the main text and supplemented here in Figs.~\ref{fig:color}

\section{Self-rotation in an atomic X system}
\begin{figure}[b]
\center{\includegraphics[width =0.4\linewidth]{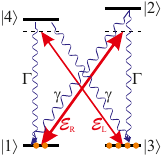}}
	\caption{Schematic diagram of energy levels with possible transitions.}
	\label{levels}
\end{figure}
We develop the theory of PSR and Eq.6 from the main text assuming X-shaped level structure as shown in Fig.~\ref{levels}. The Hamiltonian of this system under the rotating-wave approximation is as follows:
\begin{equation}
	\hat H=\hbar\begin{pmatrix}
		0&-\Omega_R&0&0\\-\Omega_R^{*}&-\Delta&0&0\\0&0&0&-\Omega_L\\0&0&-\Omega_L^{*}&-\Delta
	\end{pmatrix},
\end{equation}

We solve the master equations
\label{mastereq}
\begin{align*}
    \dot\rho_{11}=&\Gamma \rho_{44} + \gamma \rho_{22} + i \Omega_R\rho_{12}^* - i \Omega_R^{*}\rho_{12} \overset{!}{=} 0;\\
    \dot\rho_{22}=&-\Gamma \rho_{22} - \gamma \rho_{22} - i \Omega_R \rho_{12}^* + i \Omega_R^{*} \rho_{12} \overset{!}{=} 0;\\
    \dot\rho_{33}=&\Gamma \rho_{22} + \gamma \rho_{44} + i \Omega_L \rho_{34}^* - i \Omega_L^{*} \rho_{34} \overset{!}{=} 0;\\
    \dot\rho_{44}=&-\Gamma \rho_{44} - \gamma \rho_{44} - i \Omega_L \rho_{34}^* + i \Omega_L^{*} \rho_{34} \overset{!}{=} 0;\\
    \dot\rho_{12}=&- \Big(\dfrac{\gamma + \Gamma}{2} +i \Delta\Big) \rho_{12} - i \Omega_R (\rho_{11} - \rho_{22}) \overset{!}{=} 0;\\
    \dot\rho_{34}=&- \Big(\dfrac{\gamma + \Gamma}{2}  + i \Delta\Big) \rho_{34} - i \Omega_L (\rho_{33} - \rho_{44}) \overset{!}{=} 0;\\
    \dot\rho_{24}=&  i \Omega_R^* \rho_{14} - i \Omega_L \rho_{23} \overset{!}{=} 0;\\
    \dot\rho_{23}=&  i \Delta \rho_{23} + i \Omega_R^* \rho_{13} - i \Omega_L^* \rho_{24} \overset{!}{=} 0;\\
    \dot\rho_{14}=&- i \Delta \rho_{14} + i \Omega_R \rho_{24} - i \Omega_L \rho_{13} \overset{!}{=} 0;\\
    \dot\rho_{13}=&  i \Omega_R \rho_{23} - i \Omega_L^{*} \rho_{14} \overset{!}{=} 0
\end{align*}
for a steady state under the condition ${\rm Tr}(\rho_{11} + \rho_{22} + \rho_{33} + \rho_{44}) = 1$ with $\Omega_{1, 2} = \dfrac{d_{R, L} \mathcal{E}_{R, L}}{\hbar}$.  We look for the susceptibilities $\chi_{R,L}=N{\rm Tr}\hat\rho\hat d_{R,L}$ for the right and left circular polarizations, which are proportional to the mean values of the dipole moment operators
\begin{equation}
 \hat{d}_R =
 \begin{pmatrix}
  0 & d & 0 & 0 \\
  d & 0 & 0 & 0 \\
  0 & 0 & 0 & 0 \\
  0 & 0 & 0 & 0
\end{pmatrix} 
;\quad  \hat{d}_{L} =
 \begin{pmatrix}
  0 & 0 & 0 & 0 \\
  0 &  0 & 0 & 0 \\
  0 & 0 & 0 & d \\
  0 & 0 & d & 0
\end{pmatrix},
\end{equation}
yielding the refractive indices for the two polarizations:
\begin{align}
n_{R} - 1 &\approx \dfrac{1}{2} \chi_{R} =N \dfrac{{\rm Re}(d_{R} \rho_{12})}{\mathcal{E}_{R}} \\ \nonumber&\sim 
N\dfrac{ \delta}{\left(1+4 \delta^2\right)\dfrac{I_R}{I_L}+\left(1+4 \delta^2+4 \dfrac{I_R}{I_{\rm sat}}\right)}\\
n_{L}-1 &\approx \dfrac{1}{2} \chi_{L}= N \dfrac{{\rm Re}(d_{L} \rho_{34})}{E_{L}} \\& \nonumber\sim 
N\dfrac{ \delta }{\left(1+4 \delta^2\right)+\left(1+4 \delta^2+4 \dfrac{I_R}{I_{\rm sat}}\right)\dfrac{I_L}{I_R}},
\end{align}
where $N$ is the number density, $\delta = \dfrac{\Delta}{\Gamma + \gamma}$
and $I_{\rm sat\,} = \dfrac{c \epsilon_0 \hbar^2 (\Gamma + \gamma)^2}{4 \pi d^2}$.
This gives the self-rotation angle
\begin{align}
     \varphi(\varepsilon)& = 
     - \dfrac{1}{2} k l (n_R - n_L) \\ &\sim\nonumber
     \dfrac{ \delta \sin(2 \varepsilon)}{ (2 + 8 \delta^2) + I/I_{\rm sat} +  \cos(4 \varepsilon)I/I_{\rm sat}}
\end{align}

\section{Deriving Eq.~7 from main text and Eq.~\ref{eq:selfrot_formula}}
	\label{derivation}
	Let us define $\mathcal{E}_{V} \equiv V e^{i\beta}$, $\mathcal{E}_H \equiv H$ (we recall that the phase reference is  such that $\mathcal{E}_H\in\mathbb{R}$). The polarization state $\begin{pmatrix}
		\mathcal{E}_{H}\\
		\mathcal{E}_{V}
	\end{pmatrix}$ can be decomposed in the diagonal basis:
	\begin{equation}
		\begin{pmatrix}
			H\\
			V e^{i \beta}
		\end{pmatrix} 
		= 
		\dfrac{\mathcal{E}_+}{\sqrt{2}}
		\begin{pmatrix}
			1\\
			1
		\end{pmatrix}
		+
		\dfrac{\mathcal{E}_-}{\sqrt{2}}
		\begin{pmatrix}
			1\\
			-1
		\end{pmatrix},
	\end{equation}
from which we find $I_\pm\equiv|\mathcal{E}_\pm|^2 =   \dfrac{1}{2}((H \pm V \cos \beta)^2 + V^2\sin ^2 \beta)$,
and therefore	
\begin{equation}
		\cos \beta = \dfrac{|\mathcal{E}_+|^2 - |\mathcal{E}_-|^2}{2 H V}.
		\label{cos}
	\end{equation}
	The angle $\varphi$ between the major axis of the polarization ellipse  $(\mathcal{E}_H, \mathcal{E}_{V})$ and horizontal is given by \cite{peatross2011physics}
	\begin{align*}
			\varphi &= \dfrac{1}{2}\mathrm{arctan}\Big(\dfrac{2 H V \cos \beta}{H^2 - V^2}\Big)\\& = 
			\dfrac{1}{2}\mathrm{arctan}\Big(\dfrac{|\mathcal{E}_+|^2 - |\mathcal{E}_-|^2}{|\mathcal{E}_+|^2 + |\mathcal{E}_-|^2 - 2 V^2}\Big)  \\&=
			\dfrac{1}{2}\mathrm{arctan}\Big(\dfrac{I_+ - I_-}{I_+ + I_- - 2 I_V}\Big), 
		\end{align*}
where we used Eq.~\ref{cos} and that $|H|^2 + |V|^2 = |\mathcal{E}_+|^2 + |\mathcal{E}_-|^2$.
	
To derive Eq. 7 from the main text, we use the circular basis:	\begin{equation}
\begin{pmatrix}
	\mathcal{E}_{H}\\
	\mathcal{E}_{V}
\end{pmatrix}=
\begin{pmatrix}
	\mathcal{E}_{H}\\
	{\rm Re}\mathcal{E}_{V}+i	{\rm Im}\mathcal{E}_{V}
\end{pmatrix}
		= 
		\dfrac{\mathcal{E}_R}{\sqrt{2}}
		\begin{pmatrix}
			1\\
			i
		\end{pmatrix}
		+
		\dfrac{\mathcal{E}_L}{\sqrt{2}}
		\begin{pmatrix}
			1\\
			-i
		\end{pmatrix},
	\end{equation}
	which gives $I_R =|\mathcal{E}_R|^2 =  \dfrac{1}{2}((\mathcal{E}_{H} - {\rm Im}\mathcal{E}_{V})^2 +	({\rm Re}\mathcal{E}_{V})^2)$, 
	$I_L =|\mathcal{E}_L|^2 =  \dfrac{1}{2}((\mathcal{E}_{H} +{\rm Im}\mathcal{E}_{V})^2 +	({\rm Re}\mathcal{E}_{V})^2)$.
Hence
    \begin{equation}
    \begin{aligned}
        \dfrac{{\rm Im}\mathcal{E}_{V}}{\mathcal{E}_{H}} &= \dfrac{I_L - I_R}{2 I_H},\\
        \dfrac{{\rm Re}\mathcal{E}_{V}}{\mathcal{E}_{H}} = \dfrac{\sqrt{I_V - ({\rm Im}\mathcal{E}_{V})^2}}{\mathcal{E}_{H}} &= \dfrac{\sqrt{4 I_V I_H - (I_L-I_R)^2}}{2 I_H}.
    \end{aligned}
	\end{equation}

\section{Full experimental setup}
\label{suppfullsetup}

\begin{figure*}[ht!]
\includegraphics[width = 0.7\linewidth]{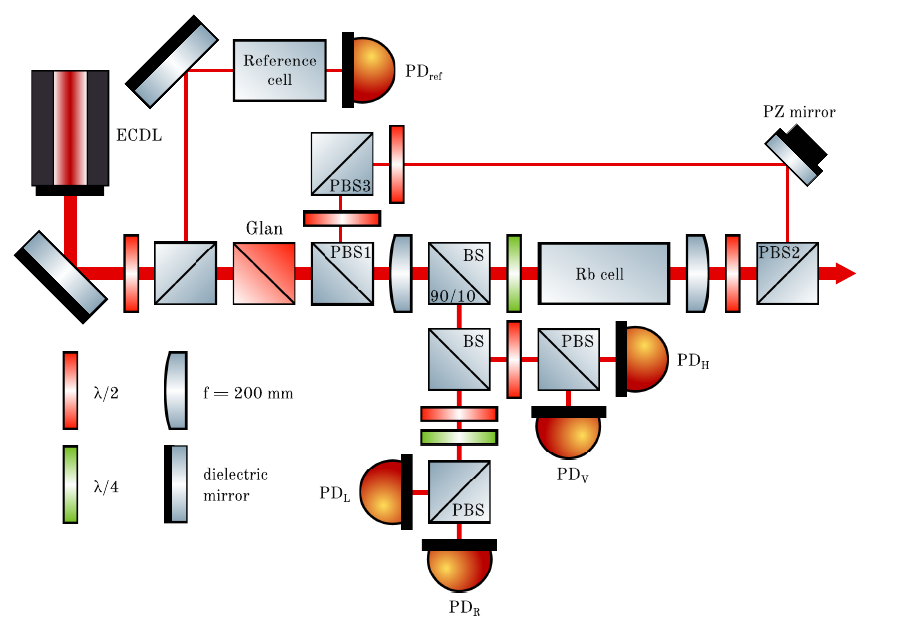}
		\caption{Full experimental setup.}
        \label{resonator setup}
	\end{figure*} 
    
We use an additional rubidium cell as a reference to determine the laser frequency using absorption spectroscopy.

Waveplates before and after the rubidium cell are needed to compensate for the birefringence of the cell's windows. To compensate for the birefringence of the cell windows and other imperfections of the setup, the polarization of the input beam is fine-tuned by these waveplates
to eliminate the cavity resonance peaks when the laser is detuned from the atomic line. 

The cavity is aligned to maximize the mode matching between the horizontally polarized light returning from PBS$_3$ and the horizontal pump by observing interference between the two fields.  We achieve the maximum visibility of $V=0.96 \pm 0.01$, which corresponds to a mode-matching efficiency of $V^2 \approx 0.92$. 

Resonator losses are controlled by rotating the $\lambda/2$ plate between PBS$_1$ and PBS$_3$ in the resonator loop (Fig.~\ref{resonator setup}).

\vspace{-0.5cm}

\end{document}